\documentclass[aps,prl,twocolumn,showpacs,superscriptaddress, twocolumn, amssymb,showpacs,floatfix,nofootinbib,tighten]{revtex4-2}
\usepackage{amssymb,amsmath,graphicx,color,microtype}
\usepackage{graphicx}
\usepackage{txfonts}
\usepackage{xcolor}
\usepackage{epsf}
\usepackage{epstopdf}
\usepackage{amssymb}
\newcommand{\nc}{\newcommand}
\newcommand\be{\begin{equation}}
\newcommand\ee{\end{equation}}

\usepackage{float}

\nc{\e}{{\bf{e}}}
\nc{\kk}{{\bf{k}}}
\nc{\pp}{{\bf{p}}}

\nc{\bfk}{{\bf{k}}}
\nc{\bfx}{{\bf{x}}}
\nc{\bfp}{{\bf{p}}}

\nc{\eH}{{\epsilon_H}}
\nc{\calP}{{\cal P}}
\nc{\im}{{ \mathrm{Im} } }

\def\apj{Astrophysical Journal}
\def\mnras{Monthly Notes of Royal Astronomical Society}
\def\apjl{Astrophysical Journal Letters}

\def\araa{Annual Review of Astronomy \& Astrophysics}

\def\aap{Astronomy and Astrophysics}     
\def\prd{Phyical Review D}
\def\prl{Phyics Review Letter}

\begin{document}

\author{Tom~Broadhurst}
\affiliation{Department of Theoretical Physics, University of the Basque Country UPV-EHU, 48040 Bilbao, Spain}
\affiliation{Donostia International Physics Center (DIPC), 20018 Donostia, The Basque Country}
\affiliation{IKERBASQUE, Basque Foundation for Science, Alameda Urquijo, 36-5 48008 Bilbao, Spain}

\author{Jose~M. Diego}
\affiliation{Instituto de F\'isica de Cantabria, CSIC-Universidad de Cantabria, E-39005 Santander, Spain}

\author{George~F.~Smoot}
\affiliation{ IAS TT \& WF Chao Foundation Professor, IAS, Hong Kong University of Science and Technology,
Clear Water Bay, Kowloon, 999077 Hong Kong, China}
\affiliation{Paris Centre for Cosmological Physics, Universit\'{e} de Paris, CNRS,  Astroparticule et Cosmologie, F-75013 Paris, France A, 10 rue Alice Domon et Leonie Duquet,
75205 Paris CEDEX 13, France}
\affiliation{Physics Department and Lawrence Berkeley National Laboratory, University of California, Berkeley,
94720 CA, USA }

\title{Interpreting LIGO/Virgo ``Mass-Gap" events as lensed Neutron Star-Black Hole binaries.}






\begin{abstract}

    Gravitational lensing allows the detection of binary black holes (BBH) at cosmological distances  with chirp masses that appear to be enhanced by $1+z$ in the range $1<z<4$, in good agreement with the reported BBH masses. We propose this effect also accounts for the puzzling ``mass gap" events (MG) newly reported by LIGO/Virgo, as distant, lensed NSBH events with $1<z<4$. The fitted mass of the neutron star member becomes $(1+z)\times 1.4M_\odot$, and is therefore misclassified as a low mass black hole. In this way, we derive a redshift of $z\simeq 3.5$ and $z\simeq 1.0$ for two newly reported ``mass asymmetric" events GW190412 \& GW190814, by interpreting them as lensed NSBH events, comprising a stellar mass black hole and neutron star. Over the past year an additional 31 BBH events and 5 MG events have been reported with high probability ($>95\%$), from which we infer a factor $\simeq 5$ higher intrinsic rate of NSBH events than BBH events, reflecting a higher proportion of neutron stars formed by early star formation. We predict a distinctive locus for lensed NSBH events in the observed binary mass plane, spanning $1<z<4$ with a narrow mass ratio, $q \simeq 0.2$, that can be readily tested when the waveform data are unlocked. All such events may show disrupted NS emission and are worthy of prompt follow-up as the high lensing magnification means EM detections are not prohibitive despite the high redshifts that we predict. Such lensed NSBH events provide an exciting prospect of directly charting the history of coalescing binaries via the cosmological redshift of their waveforms, determined relative to the characteristic mass of the neutron star member.


  \end{abstract}


\maketitle



We have proposed that most LIGO/Virgo\cite{Abbott1+2} reported binary black hole events (BBH) occur at high redshift, $1<z<4$, and are comprised of typical stellar mass ($\simeq10M_\odot$) black holes\cite{Rem}, that are highly magnified by gravitational lensing from intervening galaxies \citep{BDS1,BDS2,BDS3}. Lensing boosts the waveform amplitude allowing distant events to be detected, with a predicted mean redshift of $z\simeq 2$, thereby enhancing the observed chirp mass by a factor $1+z$ and so the characteristic stellar black holes mass $\simeq10M_\odot$ should appear at $\simeq 30M_\odot$, which in good agreement with the  reported events from O1 \& O2 where a peak at high mass is now evident \citep{BDS3}. Magnifications of typically $50-400$ are required, which can be achieved given the small sizes of GW sources when projected  close to lensing caustics. Such large magnifications are now known for individual lensed stars in high-z galaxies near the Einstein ring of lensing clusters with magnifications of over 5000 reported, \cite{Kelly,Rodney,Chen,Flashlights}. These discoveries have limited the proportion of dark matter in primordial black holes of stellar mass to $< 5$\% as only modest micro lensing is seen, consistent with the projected density of stars visible in the lensing clusters\cite{Diego2017, Oguri}. 

Our lens model also accounts for the close proportionality now evident between the reported BBH component masses \cite{BDS3}, as expected for black holes drawn from the relatively narrow mass function of stellar mass black holes in the Galaxy, which peaks at $8M_\odot$, with no black hole known below $5M_\odot$, nor above $20M_\odot$\cite{Rem,El-Badry,Stanway}. This narrow mass distribution, when lensed, appears spread out in the observed mass due to the large spread of source redshifts, and this prediction is in good agreement with the mass plane distribution of the component masses reported to date \cite{BDS3}.

The lack of spin reported for most BBH events \cite{Farr,Miller} taken together with our lensing explanation for the majority of reported BBH events may point most naturally to a dynamical binary capture origin for most BBH events, an idea that has the attraction of being long been proposed for stellar mass black hole binaries formed in dense globular star clusters\cite{Sig}. N-body simulations in this context have become more accurate, revealing  that the stellar black holes rapidly sink by dynamical friction against the numerous lower mass stars, into the dense cores of globular clusters, that predicts early binary capture of stellar mass black holes \cite{Sig,Banerjee2006}, and may also account for the enhanced abundance of close X-ray binaries in globular clusters \cite{Banerjee2006,Morscher}. Perturbing effects in these dense cores, including 3-body interactions may generate a "factory" of gravitational wave emission, vividly described as a ``mosh pit" of frenetic encounters \cite{Rodriguez2020}. A significant proportion of binaries may be retained indefinitely in the core and at larger radius \cite{Sig,Banerjee2010} or ejected from the cluster, depending on cluster mass, the details of core formation and the initial incidence of high mass binaries \cite{Breen2013,Morscher}. Individual black hole detection has recently been claimed in a nearby GC \cite{NGC3201} and full retention of stellar black holes has been advocated to explain the relatively dark core in the massive star cluster ${\rm \omega}$ Centauri \cite{Zocchi}.

 Time scale predictions for coalescence of such binaries require precise orbit hardening calculations that have been explored to date for globular clusters below $< 10^{6}M_{\odot}$ and typically without the full relativistic losses required to accurately describe close encounters between black holes and neutron stars. Careful Pre-LIGO simulations (based on GPU) predicted most BBH binaries would merge within 1~Gyr \cite{Banerjee2010}, with none expected today, but since the discovery of the apparently nearby, high mass BBH events, calculations in the context of dense star clusters have examined protracted BH merging to provide late era coalescence of black holes \cite{Rodriguez2018}. However, if we are correct in identifying an important role for lensing, then it may well be the long predicted early ``frenetic" GW production that is actually being witnessed by LIGO/Virgo. Interestingly the redshift range of the events that we have deduced for the majority of the BBH events, $1<z<4$, corresponds to the formation era of metal rich globular clusters\cite{BDS1,BDS3}, abundant in massive galaxies, and for which accurate age dating implies a formation era covering $1<z<4$ \cite{Forbes}. 
 
 Empirically we need not be concerned here with any particular formation scenario, but simply with the viability of lensing in relation to the newly recognised class of MG events. Firstly, it is harder to detect NSBH events than BBH events as the GW signal scales nearly linearly with the chirp mass, $M_{ch}^{5/6}$, which is $\sim 2.6$ times smaller for NSBH events than BBH events with a fixed BH member mass of $8M_\odot$, typical of stellar mass black holes. 
The signal-to-noise, $\rho$ at the detector is
\begin{equation}
\rho\propto \sqrt{\mu}M_{ch}^{5/6}/d_L .
\end{equation}
so this factor of 2.6 must be offset by a larger relative magnification $\mu = 2.6^2 = 6.76$, for NSBH events compared to BBH, and since the probability of being lensed scales as $\tau(>\mu) \propto \mu^{-2}$, universally for fold caustics, then we expect the relative proportion of magnified NSBH to BBH events above the detection limit to be roughly $1:45$ in favour of BBH events. Hence, a few lensed NSBH events may be anticipated for O3 given that neutron stars remnants should outnumber stellar black holes in star clusters, for a standard initial stellar mass function\cite{Banerjee2010}.

Such distant, lensed NSBH events we now argue may account for the mass gap events. Although the observed masses of these events are still not released, we may reasonably infer that one member of a lensed MG event is a neutron star with $1<z<2.5$, as then the observed neutron star mass would fall in the so called ``mass gap" defined by the lack of low mass black holes below $\simeq 5M_\odot$ in the Galaxy.  So a mass gap event is defined as $3M_\odot<M_{gap}<5M_\odot$ by LVT for classification, where the lower BH limit of $3M_\odot$ is imposed because binary neutron stars (BNS) should result in a BH remnants of about this lower mass and encouraged by the progenitor mass of the BNS event GW170817\cite{Abbot_BNS} of $2.7M_\odot$.
\footnote{Conceivably, the definition of MG could include examples of redshifted BNS, but these would have to be very highly magnified to be detected as the intrinsic chirp mass is only $1.2M_\odot$, i.e the proportion of redshifted BNS is expected only $< 10^{-2}$ 
level relative to BBH events, depending on how many of the 5 reported MG events fall in this category, requiring a much higher formation rate for detection. We do not consider this possibility further here in the absence of mass information for O3 but if pairs of MG events of near identical mass (dotted line in Figure 2) are detected, then in the lensing context we would interpret them  as BNS events at $z>1$, redshifted into the mass gap with very high magnifications $\mu \simeq 3000$ (see Eqn. 1).}  

There are 5 events classified by LIGO/Virgo as ``mass gap" events in the past year (period O3) with high probability, $\ge 95\%$. To this we add GW190414 observed in O3 and classified as BBH, because this event has a reported large mass ratio of $\simeq 3.5$ \cite{Abbot_highrat}, compared with most other BBH events which are tightly dispersed around a lower mass ratio of $1.45\pm0.07$, as highlighted in \cite{BDS3}.  The most recently reported mass-assymetric event GW190814\cite{Abbot_last} also qualifies as a lensed NSBH event from its high mass ratio, falling at the lower redshift end of our predictions, $z\simeq 1.0$. These 7 possible high redshift NSBH events can be compared with the 31 probable BBH events ($> 95\%$) reported in the same O3 period, shown in Figure~1. 

We make a detailed lens model calculation, including the distribution of magnifications and the frequency response of LIGO, making the same calculation as our previous BBH event predictions but with a neutron star instead of a black hole for lensed NSBH events. We assume for simplicity an exponential declining event rate with a timescale of 1 Gyr for NSBH binaries \citep[see Fig.3 in][]{Diego2020}, equal to the evolution we found for lensed BBH events \citep{BDS3}, and so we need add only one new free parameter for the normalization of the intrinsic rate of NSBH events relative to BBH events. We also adopt a small empirically based Gaussian dispersion for the NS masses of $0.3M_\odot$, centered on $1.4M_\odot$. The black hole masses for the NSBH events are drawn from the observed log-normal mass distribution of stellar mass black holes in our Galaxy\cite{Rem,El-Badry,Stanway,Simon-Diaz}, that peaks at $8M_\odot$, as in our previous BBH lensing work\cite{BDS3}. We adjust the relative rate parameter to reproduce the proportion of NSBH, as shown in Figure~1, finding that the observed ratio of $6/31$ for NSBH to BBH events that we find here implied to an intrinsic rate of NSBH events that is about $\simeq 5$ times higher than for BBH events. This is in agreement with our simple scaling estimate above, after allowance for the greater sensitivity at the higher frequency NSBH events that we include in our detailed lensing calculation, compared to the lower frequencies of BBH events, with relatively large chirp masses that LIGO/Virgo was not designed to detect. 

\begin{figure}[ht]
 \vspace{-1pt}
\includegraphics[width=0.5\textwidth]{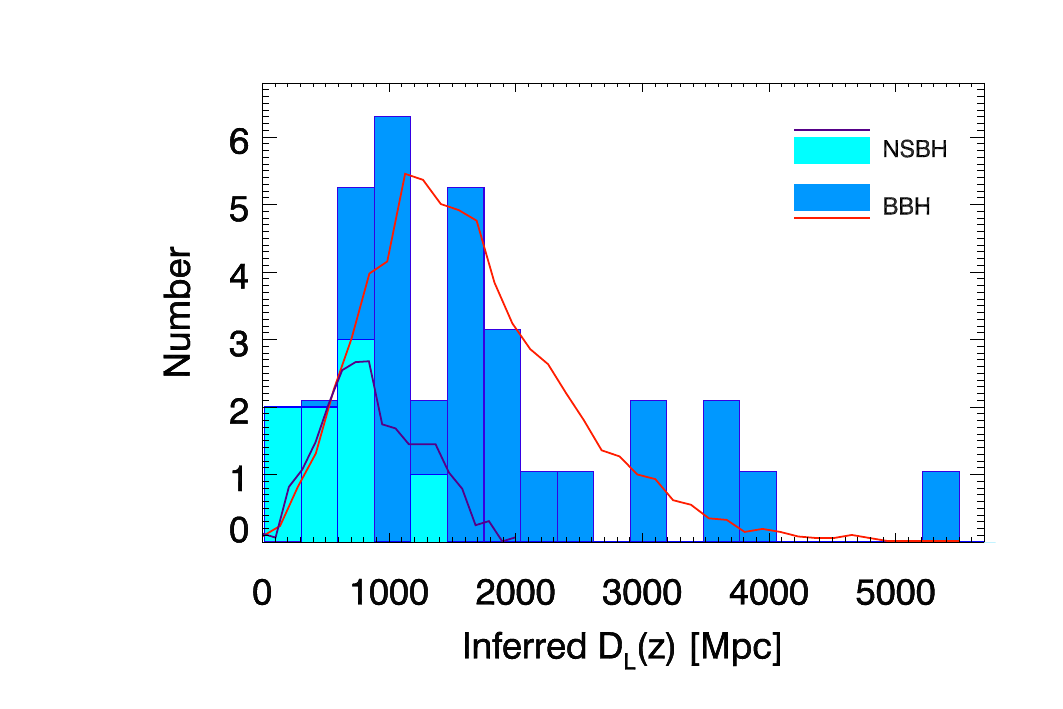}
\caption{{\bf Comparison of the reported inferred distances} for all events over the past year (period O3) with our gravitational lensing predictions for BBH events (red curve) and for NSBH events (blue curve). These predictions are compared with the events 
classified by LVT as BBH (blue) and as either NSBH or MG (cyan). The high redshift of our lensed NSBH predicted events means that most events will be classified as MG because the neutron star component fitted mass will mostly lie in the range $3-5M_\odot$. 
The intrinsic rate of NSBH events is a factor of $\simeq 5$ higher than for BBH when corrected for the selection effects which favour the detection of BBH events because of their larger chirp masses. In the context of our lensing solution, the smaller inferred distance for the NSBH reflects the relatively higher mean magnification required to detect NSBH events above the detection limit, so they appear to to be relatively nearby, closer than the inferred distances of the BBH events.
The predicted shape of these distance distributions requires no new parameters, thereby verifying our simple lens model predictions\cite{BDS1,BDS3}.} 
\end{figure}

\begin{figure}[ht]
 \vspace{-1pt}
  \includegraphics[width=0.5\textwidth]{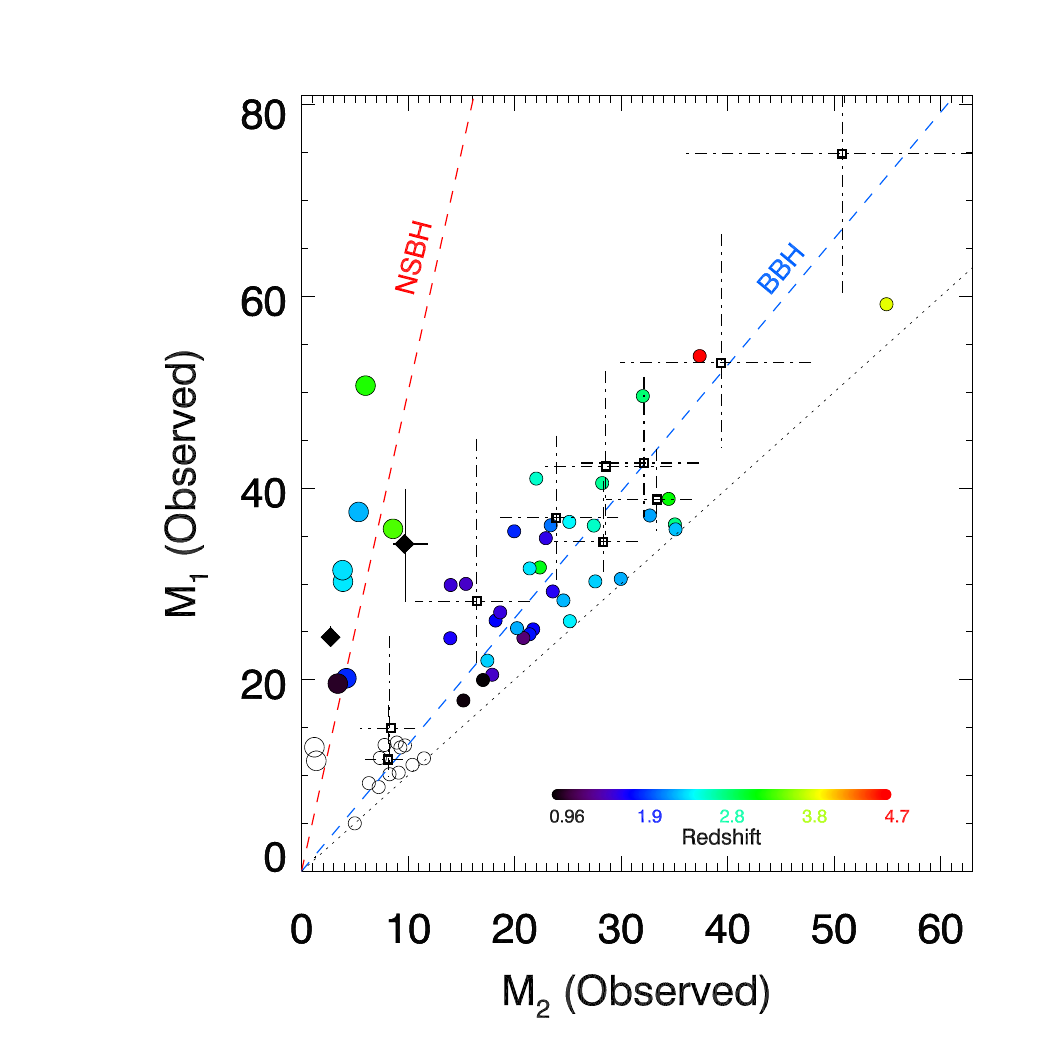}
\caption{{\bf Binary Mass Component plane.} The large circles show the predicted distribution of lensed NSBH events as a function of observed masses, colour coded by their true redshift. The observed neutron star masses, $M_2$, mostly fall within the ``mass gap", $3M_\odot<M_{gap}<5M_\odot$, simply because the waveform of the orbiting neutron star is stretched by $1+z$ in the redshift range, $1<z<4$, accessed by lensing. Large open circles indicate unlensed NSBH events predicted by our model at low redshift. The NSBH events are tightly distributed about the red dashed line indicating the mean mass ratio of $q\simeq 0.2$ based on stellar remnants in our Galaxy. The apparent mass correlation predicted here is caused by the large cosmological redshifts of the magnified events and is not an intrinsic mass relation. The same behaviour is also predicted for BBH events but with a more equal mass ratio, $q\simeq 0.7$, (blue line) based on the mass distribution of black holes in our Galaxy and agrees well with the distinctive distribution of the BBH data points for BBH events reported to date for periods O1 \& O2, shown as dashed points with error bars. Also plotted are the new mass-asymmetric events GW190412 and GW190814, as filled black diamond data points (note the errorbars are smaller than the diamond symbol for GW190814), which lie close to our predicted NSBH locus, bracketing the mass gap at the high redshift and low redshift ends.}
\label{r0-value2}
\end{figure}

\begin{figure}[ht]
 \vspace{-1pt}
  \includegraphics[width=0.5\textwidth]{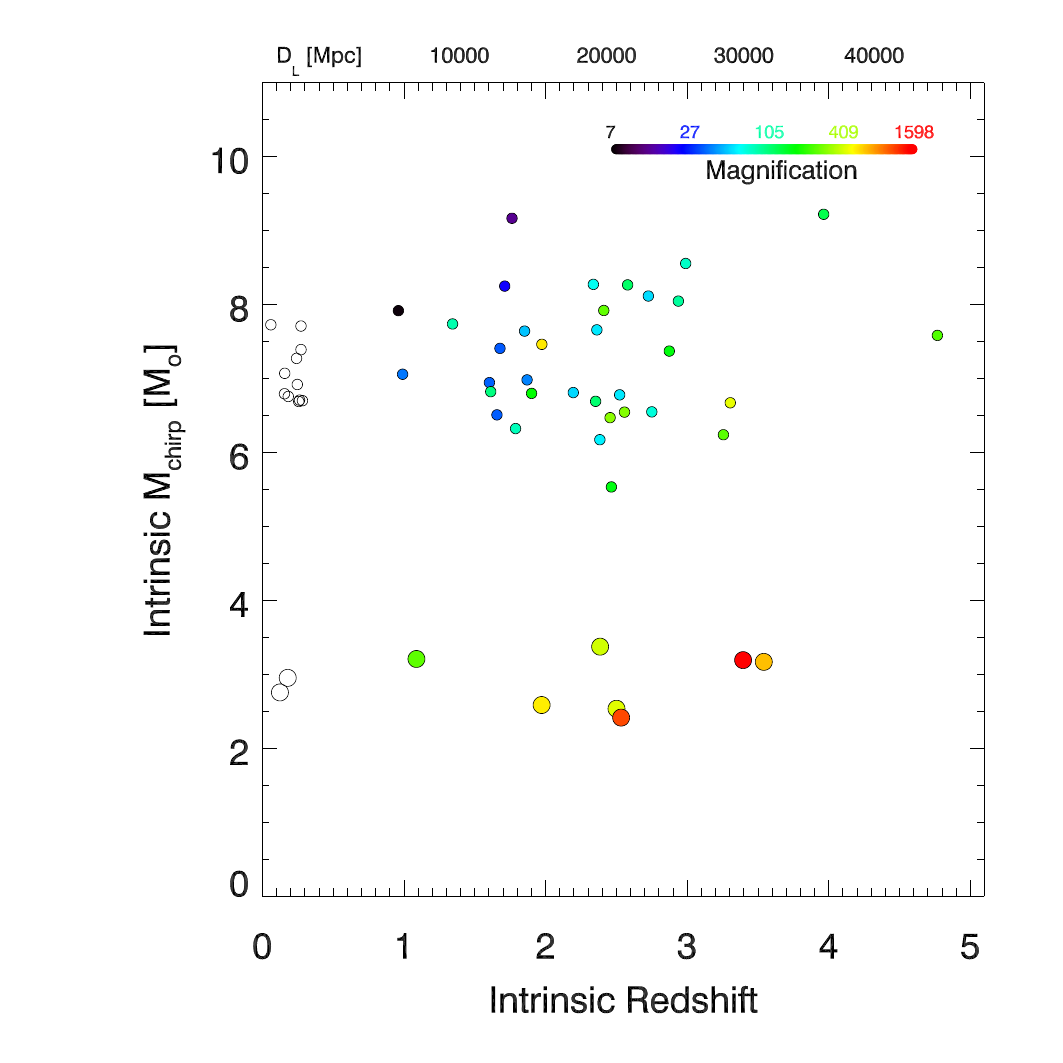}
  \vspace{-1pt}

\caption{{\bf Intrinsic Chirp masses predicted for one year of LIGO/Virgo observations} for NSBH and BBH events, corresponding to the predicted events of our lens model shown in the binary mass plane of Figure~2. These are colour coded by magnification, indicating that the lensed NSBH events (large colored circles) are typically more magnified than the lensed BBH events (small colored circles), as required to offset the lower chirp masses of the NSBH events, and hence, a lower probability of finding higher magnification which means fewer NSBH events are predicted compared to BBH events. Open circles show the unlensed events that we predict, as in Figure~2.}
\label{r0-value2}
\end{figure}

We can now make a consistency check of our lensing model using the reported {\it inferred} distances in Figure~2 for the O3 events, by converting our predicted signal strength to distance as if no lensing correction is made by the observer. Our predicted distance distribution is very similar to the NSBH events, as can be seen in Figure~1, for which the mean reported distance is $\simeq$ 780Mpc, compared to 980Mpc for the model. Also shown in Figure~1 is the good agreement for the newly reported BBH events from period O3, for which the mean reported distance is $\simeq$ 1730Mpc, which compares well with our predicted  mean depth of 1980Mpc. This smaller depth for NSBH compared to BBH events, in the absence of any lensing correction, simply reflects the higher mean magnification required to detect NSBH events than BBH events to compensate for the inherently smaller chirp masses of NSBH events (see Eqn. 1) for which the mean predicted magnification is $\mu=530$, compared to $\mu=145$ for BBH events detectable above the current sensitivity limit for period O3. The real distances for the NSBH events are far greater spanning the same source redshift range as the lensed BBH events, $1<z<4$, as can be seen in Figure~3 for the lensed NSBH and BBH events that are also shown in Figures 1\&2.

The agreement of the inferred distances of NSBH and BBH events and our simulation of lensed events, shown in Figure~1, is a satisfying independent consistency check of our lensing interpretation for the MG events as their distances were not used in constructing the lens model. This agreement verifies our adoption of the known masses of Galactic neutron stars and stellar black holes. A further more direct consistency check can be made once the binary mass estimates are released. This NSBH interpretation of MG events makes a clear prediction that the lensed NSBH events should lie along a distinctive locus in the binary mass plane, as shown in Figure~2. This follows simply from the inherently narrow spread in the ratio of stellar mass to neutron star masses in NSBH, which has a mean mass ratio of $q \simeq 0.2$, as shown in Figure~2, based on the observed masses of stellar black holes and neutron stars. 

In addition to the lensed NSBH events, our model predicts a minority of detectable events that are unlensed, as shown in the mass plane in Figure 3 (large open circles). These would be classified as NSBH events by the LIGO/Virgo team (LVT) as the redshifted mass of the neutron star is $< 3M_\odot$, as can be seen in Figure~2. In fact 2 events are classified at NSBH in period O3. This is a similar proportion of unlensed events as we predicted for BBH events\cite{BDS3} that we have used to define our rate of evolution in our previous work, where for periods O1 and 02 we have interpreted 2 of the reported BBH events with small chirp masses $\simeq 10M_\odot$, as unlensed BBH events, out of a total of 10 BBH reported, the rest of which have significantly larger chirp masses consistent with being lensed \cite{BDS3}. Note, the proportion of unlensed BBH events we predict with this model for period O3 is a little higher than the earlier periods 25\% compared to 15\% respectively, as shown in Figure~3 due to the increased current sensitivity that translates into larger volume accessible for unlensed detections.

A distinct prediction of our lensing model is that lensed NSBH events should be spread along the narrow locus visible in Figure~2, spanning the wide redshift range we predict for lensed NSBH events and all of these events should be ``mass-asymmetric", with a mass ratio of $q\simeq 0.2$, because of the inherently narrow spread in NS to BH mass ratio expected for such stellar remnants. This implies lensed NSBH events should be readily distinguishable from BBH events in the mass plane because the mass ratio of black holes is observed to be more equal, $q\simeq 0.7$, with a clear correlation, as shown in Figure~2 (dashed data points). This apparent mass correlation for BBH was readily understood in our lens model for pairs of BH drawn randomly from the relatively narrow Galactic BH distribution that we have found reproduces the BBH data very well \cite{BDS3} and is shown again in Figure~2 (but for O3 sensitivities), where we compare our BBH and NSBH predictions with the available data for periods O1 and O2. Note that in neither case is the observed correlation between the binary mass components interpreted as an intrinsic mass correlation between the two mass components, but is induced by the wide range of cosmological redshift accessed for lensed events which enhances the binary mass components inferred from redshifted waveforms.

The so called black hole ``mass gap" is an empirically defined absence of black holes with masses below $5M_\odot$ evident in the Galaxy, with a lower mass limit to this mass gap at about $3M_\odot$ as NS mergers as expected to result in a BH mass near this value, and implied directly by the NS binary GW170817\cite{Abbot_BNS}. Here we have predicted most lensed NSBH should appear to contain a NS mass that is redshifted  into this mass gap, i.e. in the range, $3M_\odot < M_{gap}< 5M_\odot$, as the lighter mass component, $m_2$ in Figure~2, is redshifted into this range for $\simeq 60\%$ of our predicted NSBH events. The redshift range we predict for lensed NSBH events extends beyond the mass gap, up to a mass of $\simeq 6.5M_\odot$ for $z\simeq 4$, following the trend indicated in Figure~2 (red dashed line) and close to the recently reported ``mass-asymmetric" event GW190412 with $q\simeq 0.2$, as shown in Figure~2 (black diamond point) which we may therefore interpret as a distant lensed NSBH event.
At the lower end of this range is the previously classified NSBH event GW190812 with $q \simeq 0.11$\cite{Abbot_last}, just reported, which falls on our predicted lensed NSBH locus(black diamond point, Figure 2) consistent with being a lensed NS of $1.4M_\odot$ and a lensed stellar mass black of $10M_\odot$, typical for known black holes in our Galaxy, at a common redshift of $z\simeq 1.0$, just below the mass gap, and fully consistent with our lens model based on Galactic black hole masses.

 Our lensing model will be readily testable soon, when we can access the waveform data for O3, raising the exciting prospect of directly defining the formation history of binary events, by dividing the measured mass of neutron star in NSBH binaries by the characteristic mass of a neutron stars, to provide $1+z$. In this way, we may infer a redshift of $z \simeq 3.5$ and $z \simeq 1.0$ for the mass asymmetric events GW190414 \& GW190814 respectively, bracketing the redshift range that we predict for lensed NSBH events in the mass gap. Such redshifted NSBH events with detected EM emission from NS disruption would pose a clear contradiction between the inferred mass from the observed waveform and that of a neutron star, thereby irresistibly implying the action of lensing. The EM flux of distant lensed events would be somewhat harder to detect than local unlensed NSBH events for the same intrinsic NSBH binary chirp, as the ratio of the lensed to unlensed EM flux is given by $\mu(d_{L_{MG}}/d_{L_{NSBH}})^{2}(1+z_{_{NSBH}})/(1+z_{_{MG}})$ where $1+z$ corrects for time dilation. This amounts to about a factor of $\simeq 5$ lower flux for associated EM emission from MG events at our mean predicted redshift, $z \simeq 2.2$, and mean magnification $\mu \simeq 500$, assuming a flat K-correction, compared to unlensed NSBH events at the maximum (unlensed) limit currently detectable with LIGO/Virgo of about 600Mpc (corresponding to GW190910). 

 \section{Discussion and Conclusions}
  
  We have shown how to make sense of the puzzling new class of ``mass gap" events discovered by LIGO/Virgo as gravitationally lensed NSBH events originating at $1<z<4$. The reported distances to these MG events are well matched by our BBH based lens model with no new free parameters, simply given by the enhanced magnification required to detect lensed NSBH events given their lower chirp masses relative to BBH events.
  
   This simple explanation for MG events as distant NSBH binaries contrasts with the contradiction they pose given the empirically based absence of low mass black holes $M < 5M_\odot$ in the Galaxy that has established the so called BH mass gap. Hence, despite the black hole classification of MG events, we advocate EM follow-up sensitive enough to detect disrupted NS emission in the redshift range $1<z<4$, which may not be prohibitive thanks to the high magnifications so that the EM flux should be only a factor of $\simeq 5$ times lower than for local, unlensed NSBH events\cite{Bhattacharya}. 
   
   This lensing based interpretation is in accordance with established astrophysical data and stellar theory. Whereas, the LVT interpretation of GW events as all being at low redshifts and unlensed, requires a new class of low mass black holes to explain the MG events and the majority of BBH events must comprise high mass black holes exceeding $\> 20M_\odot$, surpassing any known stellar mass black hole in the Galaxy. Conservative lensing models of other work implicitly assume all GW events are unlensed because they adopt a shallow extended BH mass function, that has been a popular choice, resulting in small proportion of distant lensed events that are outnumbered by foreground events of similar chirp mass. However, we can now see that most of the reported BBH masses follow a peaked mass distribution for periods O1 \& O2 and we predict this peak will become very well defined in period O3, with only a small proportion of unlensed events,$\lesssim$ 15\% of low chirp mass centered on $\simeq 10M_\odot$\cite{BDS3}.
 
  Detection of multiply-lensed events should be possible at a rate of 1 in 10 BBH cases, limited by the relatively narrow, Earth rotating footprint of LIGO/Virgo\cite{BDS2}. We have highlighted a possible pair of lensed events amongst the 10 published BBH events, with consistent waveforms and positional overlap and separated by 5 days\cite{BDS2}. Repeat detection of lensed NSBH events should be easier than for BBH events as their relatively high magnifications means each pair of events should have similar amplitudes and separated in time by as little as several minutes (and as long as days) that can be detected with uniform sensitivity. In establishing the viability of repeat lensed events we have stressed that the waveforms should be compared, not just chirp masses, so that phase and polarization consistency required by lensing can be established, in addition to chirp mass agreement \cite{BDS3}. 
  It should be kept in mind that although close pairs of lensed images should be nearly equal in brightness, most usually differ by about a factor of two\cite{Treu,SNlens}. This "flux anomaly" implies the common presence of small scale structure that is rare for standard CDM but may favour dark matter comprising light bosons\cite{Schive2014} as interference on small scales significantly perturbs lensed images near the Einstein radius\cite{Chan2020}. Hence, weaker counter images may typically fall below the detection threshold of LIGO/Virgo as most events have a SNR little above the detection limit (SNR $\simeq 8 $), allowing only rare strong events (SNR $> 16$) to have a detectable counter image.
  
  In the absence of unambiguously repeated events, the role of lensing may be established statistically by our distinctive prediction that lensed NSBH events fall along a narrow locus in the mass plane, centered on a mass ratio of $q\simeq 0.2$ set by the narrow ranges of NS and stellar BH masses observed in our Galaxy. The location of an event along this locus provides a secure redshift thanks to the characteristic neutron star mass that is enhanced by $1+z$. It may also be possible to establish the origin of GW events via the stochastic GW background, with the possibility that the full monitoring data from period O3 may usefully constrain the early high event rate we predict for our near maximal lensing interpretation of most LIGO/Virgo events\cite{Mukherjee}. Another complementary proposal is the direct detection of black holes by micro-lensing from compact substructures which perturb GW waveforms \cite{Diego2020}, that would imply the presence of a macro-lensing galaxy containing the micro-lensing substructure. These distinctive lensing predictions will be readily tested when the past year of new waveform data is unlocked, providing over 3 times more events, with the exciting prospect of directly charting the history of coalescing binaries via the cosmological redshift of MG waveforms relative to the characteristic mass of the neutron star component.



\end{document}